\documentclass[a4paper,12pt]{article}
\usepackage[top=1.2in, bottom=1in, left=1in,right=1in]{geometry}
\usepackage{amssymb}
\usepackage{subfigure}
\usepackage{graphicx}
\usepackage{epsfig}
\usepackage{booktabs}
\usepackage{array}
\usepackage{multirow}
\usepackage{cite}
\usepackage{hyperref}
\begin{document}
\title{A Model Construction of Self-similarity based Double Parton Distribution Functions for proton-proton collision at LHC}
\author{ D K Choudhury and $ \mathrm{Akbari \; Jahan}^{\star} $\\ Department of Physics, Gauhati University,\\ Guwahati - 781 014, Assam, India.\\ $ {}^{\star} $Email: akbari.jahan@gmail.com}
\date{}
\maketitle
\begin{abstract}
We construct a model for double parton distribution functions (dPDFs) based on the notion of self-similarity, pursued earlier for small \textit{x} physics at HERA. The most general form of dPDFs contains total thirteen parameters to be fitted from data of proton-proton collision at LHC. It is shown that the constructed dPDF does not factorize into two single PDFs in conformity with QCD expectation, and it satisfies the condition that at the kinematic boundary $x_{1}+x_{2}=1 $ (where $x_{1}$ and $x_{2}$ are the longitudinal fractional momenta of two partons), the dPDF vanishes.\\
\textbf{Keywords \,:} Self-similarity, parton distribution functions, small \textit{x} .
\end{abstract}
\section{Introduction}
Parton Distribution Functions (PDFs) \cite{1} are the most important quantities in Quantum Chromodynamics (QCD) to study the structure function of proton in deep inelastic scattering (DIS). At LHC, on the other hand, double Parton Distribution Functions (dPDFs) \cite{2} are equally important in the study of double parton scattering cross-sections. They are the simplest distribution functions that occur in the multipartonic interactions (MPI) which are of great relevance for LHC physics as they represent a background for the search of new physics. The physics of dPDFs are extensively discussed in the recent literature \cite{3,4,5,6,7,8,9} together with the corrected formulas for the inclusive cross-section involving dPDFs \cite{10,11,12,13,14,15}.\\

In this paper, we outline the model of PDFs based on self-similarity \cite{16}, an inherent property of fractals. Relevance of these ideas in the contemporary physics of DIS was first noted by Dremin and Levtchenko \cite{17} in the early nineties where it was shown that the saturation of hadron structure function at small \textit{x} may proceed further if the highly packed regions of proton have fractal structures. However, these ideas received wider attention in 2002 when Lastovicka \cite{18} of DESY, Hamburg proposed a relevant formalism and a functional form of the structure function $F_{2} \left(x,Q^{2} \right)$ at small \textit{x} based on self-similarity. In recent years, the present authors have applied the model to deep inelastic scattering \cite{19}, longitudinal structure function \cite{20} and momentum fractions of quarks and gluons in the proton \cite{21,22}.

\section{Formalism}
\subsection{Self-similarity based Transverse Momentum Dependent Parton Density Function (TMD PDF) with one hard scale}
The self-similarity based model of the nucleon structure function proposed in Ref.\cite{18} has been designed to be valid at small Bjorken \textit{x}. The formalism is based on the imposition of self-similarity constraints to the dimensionless quark density $f_{i} \left(x,k_{t}^{2} \right)$ and relate it to the integrated density. In other words, using magnification factors $ \frac{1}{x}$ and $ \left(1+ \frac{k_{t}^{2}}{Q_{0}^{2}} \right)$, an unintegrated quark density (TMD) is given as:
\begin{eqnarray}
\log f_{i}\left( x,k_{t}^{2}\right)& = & D_{1}\log \left( \frac{1}{x} \right)\log \left(1+\frac{k_{t}^{2}}{Q_{0}^{2}}\right)+D_{2}\log \left(\frac{1}{x}\right)+ \nonumber \\
& & D_{3}\log \left( 1+\frac{k_{t}^{2}}{Q_{0}^{2}}\right)+D_{0}^{i}-\log M^{2}
\end{eqnarray}
where \textit{i} denotes a quark flavor. Here, $ D_{2}$ and $D_{3}$ are the fractal parameters; $D_{1}$ is the dimensional correlation relating the two magnification factors; while $D_{0}^{i}$ is the normalization constant. $M^{2}$ is introduced to make PDF $q_{i} \left(x,Q^{2} \right)$ as defined in Eq (2) dimensionless. Conventional integrated quark densities (PDF) $q_{i} \left(x,Q^{2} \right)$ are defined as sum over all contributions with quark virtualities smaller than that of the photon probe $Q^{2}$. Thus $f_{i} \left(x,k_{t}^{2} \right)$ has to be integrated over $k_{t}^{2}$ to obtain $q_{i} \left(x,Q^{2} \right)$.

\begin{equation}
q_{i}\left(x,Q^{2}\right) = \int\limits_{0}^{Q^{2}}\, dk_{t}^{2} \, f_{i}\left(x,k_{t}^{2}\right)
\end{equation}
As a result, the following analytical parameterization of a quark density is obtained by using Eq (2).
\begin{equation}
q_{i}\left(x,Q^{2}\right) = e^{D_{0}^{i}} \, f \left(x,Q^{2} \right)
\end{equation}
where
\begin{equation}
f \left(x,Q^{2} \right)= \frac{Q_{0}^{2}}{M^{2}} \, \frac{x^{-D_{2}}}{1+D_{3}+D_{1} \log \left(\frac{1}{x} \right)} \, \left( \left( \frac{1}{x}\right)^{D_{1} \log \left(1+ \frac{Q^{2}}{Q_{0}^{2}} \right)} \, \left(1+ \frac{Q^{2}}{Q_{0}^{2}} \right)^{D_{3}+1} -1 \right)
\end{equation}
is flavor independent and $e^{D_{0}^{i}}$ is the only flavor dependent parameter.
Using Eq (3) in the usual definition of the structure function $F_{2} \left(x,Q^{2} \right)$
\begin{equation}
F_{2}\left(x,Q^{2} \right) = x \, \sum_{i} e_{i}^{2} \left(q_{i} \left(x,Q^{2}\right)+ \bar{q_{i}}\left(x,Q^{2}\right) \right)
\end{equation}
one has
\begin{equation}
F_{2}\left(x,Q^{2}\right) = e^{D_{0}} \, x \, f \left(x,Q^{2} \right)
\end{equation}
where $ D_{0} = \sum_{i} D_{0}^{i}$.
From HERA data \cite{23,24}, Eq (4) was fitted with
\begin{eqnarray}
D_{0} & = & 0.339 \pm 0.145 \nonumber \\
D_{1} & = & 0.073\pm 0.001 \nonumber \\
D_{2} & = & 1.013\pm 0.01 \nonumber \\
D_{3} & = & -1.287\pm 0.01 \nonumber \\
Q_{0}^{2} & = & 0.062\pm 0.01 \; \mathrm{GeV}^{2}
\end{eqnarray}
in the kinematical region
\begin{equation}
6.2\, \times \, 10^{-7} \leq x \leq 10^{-2} \quad \mathrm{and} \quad 0.045 \leq Q^{2} \leq 120 \, \mathrm{GeV^{2}}
\end{equation}
We set $M^{2}=1 \, \mathrm{GeV^{2}}$.

\subsection{Self-similarity based small \textit{x} TMD PDF extrapolated to large \textit{x}}
The model of Ref.\cite{18} was tested for a limited range of small \textit{x} as noted in Eq (8). It did not take into account the large \textit{x} behavior \cite{1,25,26} of the PDF or structure function
\begin{equation}
\lim_{x\rightarrow 1} \, F_{2} \left(x,Q^{2} \right)=0
\end{equation}
which is not unexpected. The important observation which motivated and justified the use of self-similarity concept was that for $ x<0.01$; the logarithm of the derivative of the unintegrated parton distribution $ \log \left(\frac{\partial f_{i}\left(x,Q^{2} \right)}{\partial Q^{2}} \right)$ is a linear function of $ \log x $(Fig. 2.8.a of Ref.\cite{18}). The idea of self-similarity is based on the fact that at small \textit{x}, the behavior of quark density is driven by gluon emissions and splittings such that the parton distribution function at small \textit{x} and those at still smaller \textit{x} look similar (upto some magnification factor). In the opposite limit, of large \textit{x}, there is no physical reason for self-similarity and no phenomenological justification till date. In other words, extending the approach of Ref.\cite{18} to $ x>0.01$ means applying the self-similarity concept where it is not expected to work. On the other hand, it is not unreasonable to assume that the self-similarity does not terminate abruptly at $ x\approx 0.01$, but smoothly vanishes at $x=1$, the valence quark limit of proton with no trace of self-similarity at all.\\

We take this alternative point of view in the present subsection. We suggest a simple interpolating model of TMD PDF / PDF which approaches the self-similar one at $ x\rightarrow 0$ (Eq (1)), and still satisfy Eq (9) at large \textit{x}, $ x\rightarrow 1$. A plausible way of achieving it in a parameter-free way is to make a formal replacement of $ \frac{1}{x}$ factor to $ \left( \frac{1}{x}-1\right)$ in Eq (1). The former one is identified as one of the magnification factors in the self-similar model, while the later can be so interpreted only for $ \frac{1}{x} >> 1$. In such case, Eq (1) will be modified to $ \tilde{f_{i}}\left(x,k_{t}^{2} \right)$ defined as:
\begin{eqnarray}
\log \tilde{f_{i}}\left( x,k_{t}^{2}\right)& = & D_{1}\log \left( \frac{1}{x}-1 \right)\log \left(1+\frac{k_{t}^{2}}{Q_{0}^{2}}\right)+D_{2}\log \left(\frac{1}{x}-1\right)+ \nonumber \\
& & D_{3}\log \left( 1+\frac{k_{t}^{2}}{Q_{0}^{2}}\right)+D_{0}^{i}-\log M^{2}
\end{eqnarray}
This leads to the expression for PDF as
\begin{equation}
\tilde{q_{i}}\left(x,Q^{2}\right) = e^{D_{0}^{i}} \, \tilde{f} \left(x,Q^{2} \right)
\end{equation}
and the structure function
\begin{equation}
\tilde{F_{2}}\left(x,Q^{2}\right) = e^{D_{0}} \, x \, \tilde{f} \left(x,Q^{2} \right)
\end{equation}
where
\begin{small}
\begin{eqnarray}
\tilde{f}\left(x,Q^{2} \right) & = & \frac{Q_{0}^{2}}{M^{2}} \, \frac{x^{-D_{2}} \, (1-x)^{D_{2}}}{1+D_{3}+D_{1} \log \left(\frac{1}{x} \right)+ D_{1} \log (1-x)} . \nonumber \\
& & \left( \left( 1-x \right)^{D_{1} \log \left(1+ \frac{Q^{2}}{Q_{0}^{2}} \right)} x^{-D_{1} \log \left(1+ \frac{Q^{2}}{Q_{0}^{2}} \right)}  \left(1+ \frac{Q^{2}}{Q_{0}^{2}} \right)^{D_{3}+1} -1 \right)
\end{eqnarray}
\end{small}
which is flavor independent and $ D_{0} = \sum \limits_{i} D_{0}^{i}$.\\

It is desirable to discuss the relation of the parameterization (Eq (13)) with the common behavior of quark and gluon distributions obtained in the framework of perturbative QCD with standard parameterization like CTEQ \cite{27}. In the double leading logarithmic approximation, the explicit forms of single distribution functions on the parton level are also well-known \cite{28,29}. It is therefore of interest to know if these explicit perturbative parton distributions in the region of small \textit{x} are self-similar or not.\\

Using Eq (13) in Eq (11) and setting $Q^{2}=Q_{0}^{2}$, we have
\begin{equation}
x \, \tilde{q_{i}}\left(x,Q_{0}^{2}\right) = e^{D_{0}^{i}} \frac{Q_{0}^{2}}{M^{2} \, l(x)} \left\lbrace x^{1-D_{1} \log 2 - D_{2}} \left(1-x \right)^{D_{2}+D_{1} \log 2}.2^{D_{3}+1}- x^{1-D_{2}} \left(1-x \right)^{D_{2}} \right\rbrace
\end{equation}
where
\begin{equation}
l(x)=1+D_{3}+D_{1}\log \left(\frac{1}{x} \right)+D_{1}\log \left(1-x \right)
\end{equation}
The \textit{x} dependence of $ l(x) $ is due to the assumed correlation between the two magnification factors $ \frac{1}{x}$ and $ \left(1+ \frac{k_{t}^{2}}{Q_{0}^{2}} \right)$ (Eq (1)). If it is assumed to be negligible, then Eq (14) has a form similar to the canonical parameterization \cite{1,25},
\begin{equation}
x \, q_{i}\left(x,Q_{0}^{2}\right) \approx A_{0}^{i} \,  x^{A_{1}^{i}} \left(1-x \right)^{A_{2}^{i}}
\end{equation}
where the superscript \textit{i} indicates flavor dependence. If $n_{f}$ is the number of flavors for both quarks and antiquarks, then the number of parameters in Eq (16) will be $ 6n_{f}+3 $; 3 being the number of parameters for the gluon distribution.\\

In a self-similar parameterization like Eq (14), the exponents of \textit{x} and $\left(1-x \right)$ factors are flavor independent: each flavored quark can just be scaled up or down without changing the shapes in \textit{x}-plane. Thus in the self-similar quark and antiquark distributions (including gluon) (Eq (14)), total number of parameters will be $ 2n_{f}+3+4 $ (additional 4 coming from the gluon distribution), a decrease of number by $ 4n_{f}-4 $. The recent CTEQ parameterization \cite{27} has, on the other hand, a form, which is generalisation of the canonical form (Eq (16)):
\begin{equation}
x \, q_{i}\left(x,Q_{0}^{2}\right) = A_{0}^{i} \,  x^{A_{1}^{i}} \left(1-x \right)^{A_{2}^{i}} \, e^{A_{3}^{i}x} \, \left(1+e^{A_{4}^{i}}x\right)^{A_{5}^{i}}
\end{equation}
A similar 6 parameter form can also be written for gluon distribution $ xg \left(x,Q_{0}^{2} \right)$. In this case, total number of parameters for quarks, antiquarks and gluon will be $ 12n_{f}+6 $.\\

The above analysis indicates that in the absence of correlation between the magnification factors, the self-similarity based parameters (Eq (13)) has strong resemblance to the canonical parameterization (Eq (16)).\\

In Ref.\cite{27,28}, nth moment of the valence (non-singlet) and sea quark (antiquark) distributions are reported as
\begin{equation}
q^{valence} \left(n, \zeta \right)= exp \left(\nu_{0}(n).\zeta \right)
\end{equation}
\\
\begin{equation}
q^{sea} \left(n, \zeta \right)=\frac{1}{2n_{f}} \left[\frac{\nu_{0}(n)-\nu_{\_}(n)}{\nu_{+}(n)-\nu_{\_}(n)}exp\left(\nu_{+}(n).\zeta \right)+\frac{\nu_{+}(n)-\nu_{0}(n)}{\nu_{+}(n)-\nu_{\_}(n)}exp\left(\nu_{\_}(n).\zeta \right)-exp\left(\nu_{0}(n).\zeta \right)\right]
\end{equation}

Here
\begin{eqnarray}
\zeta=\frac{1}{\beta_{0}}\ln \left[\frac{\ln \left(\frac{Q^{2}}{\Lambda^{2}}\right)}{\ln \left(\frac{\mu^{2}}{\Lambda^{2}}\right)} \right] \nonumber\\
\beta_{0}=11-\frac{2}{3}n_{f}
\end{eqnarray}
and $\nu_{0}(n)$, $\nu_{\pm}(n)$ are functions of \textit{n} which are identified as anomalous dimensions and whose explicit forms are given in \cite{27}. The above equations (Eq (18) and Eq (19)) show that the nth moment of quark/antiquark of any flavor transforms like
\begin{equation}
q \left(n,Q^{2} \right) \sim \left(\frac{\ln \left(\frac{Q^{2}}{\Lambda^{2}}\right)}{\ln \left(\frac{\mu^{2}}{\Lambda^{2}}\right)} \right)^{\frac{\nu_{i}(n)}{\beta_{0}}}
\end{equation}
where $ i = 0, +, - $.\\

To see if the corresponding explicit parton distributions are self-similar, we obtain the nth moment of the self-similar PDF (Eq (3)) defined as
\begin{equation}
q \left(n,Q^{2}\right)= \int \limits_{0}^{1} dx \, x^{n-1} \, q_{i}\left(x,Q^{2} \right)
\end{equation}
and is found to transform like
\begin{equation}
q \left(n,Q^{2}\right) \sim \sum_{i=0}^{\infty} c_{i}\left[\log \left(1+\frac{Q^{2}}{Q_{0}^{2}}\right) \right]^{i}
\end{equation}

Comparing Eq (23) with Eq (21), we infer that the QCD parton distributions of Ref.\cite{28} can be considered approximately self-similar in the sense that it transforms like power of $ \ln \frac{Q^{2}}{\Lambda^{2}}$ where the exponent is identified as the anomalous dimension.

\subsection{Self-similar TMD dPDF and dPDF at small $x_{1} \, , \, x_{2}$}
In this case, one has hard scales $k_{t}^{2(1)}$ and $k_{t}^{2(2)}$ of partons carrying fractional momentum $x_{1}$ and $x_{2}$ of flavors \textit{i} and \textit{j} respectively. Corresponding to the virtuality $ Q^{2}$ of DIS, the partons will also have hard scales $ Q^{2(1)}$ and $ Q^{2(2)}$. For simplicity, we assume them to be fixed. Following Ref.\cite{18}, the TMD dPDF for partons of flavors \textit{i} and \textit{j} will then have the following basic magnification factors:
\begin{eqnarray}
M_{1} & = & \frac{1}{x_{1}} \nonumber \\
M_{2} & = & \frac{1}{x_{2}} \nonumber \\
M_{3} & = & \left(1+ \frac{k_{t}^{2(1)}}{k_{0}^{2}} \right) \nonumber \\
M_{4} & = & \left(1+ \frac{k_{t}^{2(2)}}{k_{0}^{2}} \right)
\end{eqnarray}
\\
The TMD dPDF for a pair of partons of flavors \textit{i} and \textit{j} is then given as:
\begin{eqnarray}
\log f_{ij} \left( x_{1},x_{2},k_{t}^{2(1)},k_{t}^{2(2)}\right) & = & D_{1} \log M_{1} \log M_{2} + D_{2} \log M_{1} \log M_{4} + D_{3} \log M_{2} \log M_{3} \nonumber \\
& & + D_{4} \log M_{3} \log M_{4} + D_{5} \log M_{1} \log M_{2} \log M_{3}  \nonumber \\
& & + D_{6} \log M_{1} \log M_{2} \log M_{4} + D_{7} \log M_{1} \log M_{3} \log M_{4}  \nonumber \\
& & + D_{8} \log M_{2} \log M_{3} \log M_{4} + D_{9} \log M_{1} \log M_{2} \log M_{3} \log M_{4} \nonumber \\
& & + D_{0}^{ij} - \log M^{4}
\end{eqnarray}
\\
Eq (25) has total 11 parameters, viz. 9 flavor independent fractal parameters, $ D_{1},...,D_{9}$, one flavor dependent normalization constant $D_{0}^{ij}$ and a mass scale $M^{4}$. The additional term  $\log M^{4}$ is added in Eq (25) with dimension $ (\mathrm{mass})^{4}$ so that the dPDF defined as
\begin{equation}
D_{ij} \left(x_{1},x_{2},Q^{2(1)},Q^{2(2)} \right)= \int\limits_{0}^{Q^{2(1)}}\, dk_{t}^{2(1)} \, \int\limits_{0}^{Q^{2(2)}}\, dk_{t}^{2(2)} \, f_{ij}\left(x_{1},x_{2},k_{t}^{2(1)},k_{t}^{2(2)} \right)
\end{equation}
is dimensionless.\\

TMD dPDF (Eq (25)) can be rewritten as
\begin{eqnarray}
f_{ij} \left( x_{1},x_{2},k_{t}^{2(1)},k_{t}^{2(2)}\right) & = & \frac{e^{D_{0}^{ij}}}{M^{4}} \, \left( \frac{1}{x_{1}}\right)^{D_{1} \log \frac{1}{x_{2}}} \, \left( \frac{1}{x_{1}}\right)^{D_{2} \log \left(\frac{k_{t}^{2 (2)}+k_{0}^{2}}{k_{0}^{2}} \right)} \, \left( \frac{1}{x_{2}}\right)^{D_{3} \log \left(\frac{k_{t}^{2 (1)}+k_{0}^{2}}{k_{0}^{2}} \right)} . \nonumber \\
& & \left(\frac{k_{t}^{2 (1)}+k_{0}^{2}}{k_{0}^{2}} \right)^{D_{4} \log \left(\frac{k_{t}^{2 (2)}+k_{0}^{2}}{k_{0}^{2}} \right)} \, \left( \frac{1}{x_{1}}\right)^{D_{5} \log \frac{1}{x_{2}} \log \left(\frac{k_{t}^{2 (1)}+k_{0}^{2}}{k_{0}^{2}} \right)} . \nonumber \\
& &  \left( \frac{1}{x_{1}}\right)^{D_{6} \log \frac{1}{x_{2}} \log \left(\frac{k_{t}^{2 (2)}+k_{0}^{2}}{k_{0}^{2}} \right)} \, \left( \frac{1}{x_{1}}\right)^{D_{7} \log \left(\frac{k_{t}^{2 (1)}+k_{0}^{2}}{k_{0}^{2}} \right)\log \left(\frac{k_{t}^{2 (2)}+k_{0}^{2}}{k_{0}^{2}} \right)} . \nonumber \\
& & \left( \frac{1}{x_{2}}\right)^{D_{8} \log \left(\frac{k_{t}^{2 (1)}+k_{0}^{2}}{k_{0}^{2}} \right)\log \left(\frac{k_{t}^{2 (2)}+k_{0}^{2}}{k_{0}^{2}} \right)} . \nonumber \\
& & \left(\frac{k_{t}^{2 (2)}+k_{0}^{2}}{k_{0}^{2}} \right)^{D_{9} \log \frac{1}{x_{1}} \log \frac{1}{x_{2}} \log \left(\frac{k_{t}^{2 (1)}+k_{0}^{2}}{k_{0}^{2}} \right)}
\end{eqnarray}
For $ k_{t}^{2}$ integration, it will be more convenient to write it in the form
\begin{eqnarray}
f_{ij} \left( x_{1},x_{2},k_{t}^{2(1)},k_{t}^{2(2)}\right) & = & \frac{e^{D_{0}^{ij}}}{M^{4}} \, e^{D_{1} \log \frac{1}{x_{1}} \log \frac{1}{x_{2}}} \, \left(\frac{k_{t}^{2 (2)}+k_{0}^{2}}{k_{0}^{2}} \right)^{D_{2} \log \frac{1}{x_{1}}} \, \left(\frac{k_{t}^{2 (1)}+k_{0}^{2}}{k_{0}^{2}} \right)^{D_{3} \log \frac{1}{x_{2}}}. \nonumber \\
& & \left(\frac{k_{t}^{2 (1)}+k_{0}^{2}}{k_{0}^{2}} \right)^{D_{4} \log \left(\frac{k_{t}^{2 (2)}+k_{0}^{2}}{k_{0}^{2}} \right)} \, \left(\frac{k_{t}^{2 (1)}+k_{0}^{2}}{k_{0}^{2}} \right)^{D_{5} \log \frac{1}{x_{1}} \log \frac{1}{x_{2}}} .  \nonumber \\
& & \left(\frac{k_{t}^{2 (2)}+k_{0}^{2}}{k_{0}^{2}} \right)^{D_{6} \log \frac{1}{x_{1}} \log \frac{1}{x_{2}}} \, \left(\frac{k_{t}^{2 (1)}+k_{0}^{2}}{k_{0}^{2}} \right)^{D_{7} \log \frac{1}{x_{1}} \log \left(\frac{k_{t}^{2 (2)}+k_{0}^{2}}{k_{0}^{2}} \right)} . \nonumber \\
& & \left(\frac{k_{t}^{2 (2)}+k_{0}^{2}}{k_{0}^{2}} \right)^{D_{8} \log \frac{1}{x_{2}} \log \left(\frac{k_{t}^{2 (1)}+k_{0}^{2}}{k_{0}^{2}} \right)} . \nonumber \\
& & \left(\frac{k_{t}^{2 (2)}+k_{0}^{2}}{k_{0}^{2}} \right)^{D_{9} \log \frac{1}{x_{1}} \log \frac{1}{x_{2}} \log \left(\frac{k_{t}^{2 (1)}+k_{0}^{2}}{k_{0}^{2}} \right)}
\end{eqnarray}
\\
Using the definition of dPDF (Eq (26)), one has
\begin{equation}
D_{ij} \left(x_{1},x_{2},Q^{2(1)},Q^{2(2)} \right)= \frac{e^{D_{0}^{ij}}}{M^{4}} \, \left( \frac{1}{x_{1}}\right)^{D_{1} \log \frac{1}{x_{2}}} \, . \, I \left(x_{1},x_{2},Q^{2(1)},Q^{2(2)} \right)
\end{equation}
where $ I \left(x_{1},x_{2},Q^{2(1)},Q^{2(2)} \right) $ is the double integration over $k_{t}^{2(1)}$ and $k_{t}^{2(2)}$. That is, \\

$I \left(x_{1},x_{2},Q^{2(1)},Q^{2(2)} \right) =$
\begin{eqnarray}
 & & \int\limits_{0}^{Q^{2(1)}}\, dk_{t}^{2(1)} \, \int\limits_{0}^{Q^{2(2)}}\, dk_{t}^{2(2)} \, \left(\frac{k_{t}^{2 (1)}+k_{0}^{2}}{k_{0}^{2}} \right)^ {D_{3}\log \frac{1}{x_{2}}+ D_{4} \log \left(\frac{k_{t}^{2 (2)}+k_{0}^{2}}{k_{0}^{2}} \right)+D_{5} \log \frac{1}{x_{1}} \log \frac{1}{x_{2}}+D_{7} \log \frac{1}{x_{1}} \log \left(\frac{k_{t}^{2 (2)}+k_{0}^{2}}{k_{0}^{2}} \right)}. \nonumber \\
& & \left(\frac{k_{t}^{2 (2)}+k_{0}^{2}}{k_{0}^{2}} \right)^{D_{2} \log \frac{1}{x_{1}}+ D_{6} \log \frac{1}{x_{1}} \log \frac{1}{x_{2}}+ D_{8}\log \frac{1}{x_{2}} \log \left(\frac{k_{t}^{2 (1)}+k_{0}^{2}}{k_{0}^{2}} \right)+ D_{9} \log \frac{1}{x_{1}} \log \frac{1}{x_{2}} \log \left(\frac{k_{t}^{2 (1)}+k_{0}^{2}}{k_{0}^{2}} \right)}
\end{eqnarray}
Eq (29) is our main result for self-similar dPDF at small $x_{1}$ and $x_{2}$. It contains total 12 parameters, viz. 9 fractal parameters $(D_{1}, ... ,D_{9})$, one normalization constant $D_{0}^{ij}$, one mass scale $M^{4}$ and one transverse mass cut off $k_{0}^{2}$. Before proceeding further, we note that the integration (Eq (30)) is not factorisable in $x_{1}$ and $x_{2}$. Even the multiplicative term  $ \left(\frac{1}{x_{1}} \right)^{D_{1}\log \frac{1}{x_{2}}}$ of Eq (29) is not so. Thus the usual factorisability assumption \cite{30} that a dPDF can be considered as a product of two single PDF
\begin{equation}
D_{ij} \left(x_{1},x_{2},Q^{2(1)},Q^{2(2)} \right) \equiv D_{i}\left( x_{1},Q^{2(1)} \right). D_{j}\left( x_{2},Q^{2(2)} \right)
\end{equation}
does not hold good in the present self-similarity based dPDF.\\

It is to be noted that the factorized assumption (Eq (31)) is merely a simple assumption and is not based on QCD. Its status was first discussed by Snigirev \cite{3} where it was shown that the distributions of two partons are correlated in the leading logarithmic approximation, i.e.
\begin{equation}
D_{ij} \left(x_{1},x_{2},Q^{2(1)},Q^{2(2)} \right) \neq D_{i}\left( x_{1},Q^{2(1)} \right). D_{j}\left( x_{2},Q^{2(2)} \right)
\end{equation}
It was further observed that \cite{3} even if the two parton distributions are factorised at some scale $ \mu^{2} $, then the QCD evolution violates such factorisation invariably at any different scale $ Q^{2}\neq \mu^{2} $. The first estimation at such perturbative QCD correlation was reported by Korotkikh and Snigirev \cite{4} at LHC scale $\left(\sim 100 \, \mathrm{GeV} \right)$: it is as large as $30\%$. Thus the present self-similarity based model of dPDF conform to QCD expectation of non-factorisability (Eq (32)).

\subsection{TMD dPDF and dPDF at the boundary $x_1+x_2=1$}
The standard behavior of single PDF is
\begin{equation}
\lim_{x_{1}\rightarrow 1} \, D_{i}\left(x_{1},Q^{2(1)} \right)=0
\end{equation}
and
\begin{equation}
\lim_{x_{2}\rightarrow 1} \, D_{j}\left(x_{2},Q^{2(2)} \right)=0
\end{equation}
The corresponding boundary condition of dPDF, on the other hand, is \cite{2}
\begin{equation}
\lim_{x_{1}+x_{2}\rightarrow 1} \, D_{ij}\left(x_{1},x_{2},Q^{2(1)},Q^{2(2)} \right)=0
\end{equation}
which does not conform to Eq (33) and Eq (34).\\

Thus the simple assumption of factorisability of dPDF into two PDF fails at the kinematic boundary $x_{1}+x_{2}=1$. So usually the dPDF is multiplied by a phenomenological factor $\rho_{ij}\left(x_{1},x_{2} \right)$ of the term \cite{2}
\begin{equation}
\rho_{ij}\left(x_{1},x_{2} \right) \sim \left( 1-x_{1}-x_{2} \right)^{2}.\left( 1-x_{1}\right)^{-2- \alpha}.\left( 1-x_{2}\right)^{-2- \alpha}
\end{equation}
where $ \alpha $ is zero for sea quarks and 0.5 for valence partons \cite{2}.\\

We note that the form of Eq (36) was suggested by Gaunt and Stirling \cite{5}. It was an improvement over the earlier one,
\begin{equation}
\rho_{ij}\left(x_{1},x_{2} \right) \sim \left( 1-x_{1}-x_{2} \right)^{n}
\end{equation}
to satisfy the dPDF number sum rules \cite{5}. That at \textit{x} close to 1, the dPDF should in general include the factors:
\begin{equation}
\rho_{ij}\left(x_{1},x_{2} \right) \sim \left( 1-x_{1}-x_{2} \right)^{\alpha}.\left( 1-x_{1}\right)^{\beta}.\left( 1-x_{2}\right)^{\gamma}
\end{equation}
with the exponents $\alpha,\, \beta, \, \gamma $ depending on parton types \cite{7}.\\

As described above, the notion of self-similarity for the PDF of large $x_{1}, \, x_{2}$ is not expected to hold. Unlike TMD PDF, there is also no simple parameter-free prescription for incorporating the kinematic boundary condition (Eq (35)) so that it coincides with the original definition (Eq (25)) for $x_{1},x_{2}\rightarrow 0$. A simple way of incorporating such effect is to introduce an additional factor $ M_{5}= \left(\frac{1}{x_{1}+x_{2}}-1 \right)$,which for $x_{2}=0$ and $x_{1}\rightarrow 0 $ ($x_{1}=0$ and $x_{2}\rightarrow 0 $) approaches $M_{1} (M_{2})$ of Eq (24). $ \log f_{ij} \left(x_{1},x_{2},k_{t}^{2(1)},k_{t}^{2(2)} \right)$ defined in Eq (25) then gets the form
\begin{equation}
\log \tilde{f}_{ij} \left(x_{1},x_{2},k_{t}^{2(1)},k_{t}^{2(2)} \right)=\log f_{ij} \left(x_{1},x_{2},k_{t}^{2(1)},k_{t}^{2(2)} \right)+ D_{10} \log M_{5}
\end{equation}
which results in
\begin{equation}
\tilde{f}_{ij} \left(x_{1},x_{2},k_{t}^{2(1)},k_{t}^{2(2)} \right)= f_{ij} \left(x_{1},x_{2},k_{t}^{2(1)},k_{t}^{2(2)} \right). \left(\frac{1}{x_{1}+x_{2}}-1 \right)^{D_{10}}
\end{equation}
Eq (40) is the expression for the TMD dPDF compatible with the boundary condition (Eq (35)). At small $x_{1},x_{2}$ it has self-similar basis. The corresponding dPDF expression is
\begin{equation}
\tilde{D}_{ij} \left(x_{1},x_{2},Q^{2(1)},Q^{2(2)} \right)= \frac{e^{D_{0}^{ij}}}{M^{4}} \, \left( \frac{1}{x_{1}}\right)^{D_{1} \log \frac{1}{x_{2}}} \, .\left(\frac{1- \left(x_{1}+x_{2} \right)}{x_{1}+x_{2}} \right)^{D_{10}} . \, \tilde{I} \left(x_{1},x_{2},Q^{2(1)},Q^{2(2)} \right)
\end{equation}
\\
where $\tilde{I} \left(x_{1},x_{2},Q^{2(1)},Q^{2(2)} \right)$ is the double integration over the transverse momenta $k_{t}^{2(1)}$ and $k_{t}^{2(2)}$. It has the same expression as that of $ I \left(x_{1},x_{2},Q^{2(1)},Q^{2(2)} \right)$ as given in Eq (30).\\

Eq (41) of the final expression for dPDF in the approach contains total thirteen parameters to be determined from LHC data. It is to be noted that Eq (41) does not yield Eq (29) in small $x_{1},x_{2}$ limit unlike Eq (11) which leads to Eq (3) at small \textit{x} limit. Such a smooth extrapolation is possible only if $D_{10}$ itself has $x_{1},x_{2}$ dependence such that $ \lim \limits_{x_{1},x_{2}\rightarrow 0} \, D_{10} \left(x_{1},x_{2} \right)=0$. Given the paucity of experimental data regarding dPDF, we discuss only some rough qualitative features of the model graphically in the next section.

\section{Graphical Analysis of PDFs and dPDFs}
For qualitative feature of the model, we plot in Figure 1 the PDF $q_{i}\left(x,Q^{2} \right)$ vs \textit{x} for representative values of $Q^{2}=10, 50 \, \mathrm{and}\, 100 \, \mathrm{GeV^{2}}$ of HERA range using Eq (3) (dashed lines) and Eq (11) (solid lines). It shows the qualitative difference between the original model of Ref.\cite{18} and the smooth extrapolation for large \textit{x}.
\\
\begin{figure}[h]
\centering
\mbox{\subfigure{\includegraphics[scale=.56]{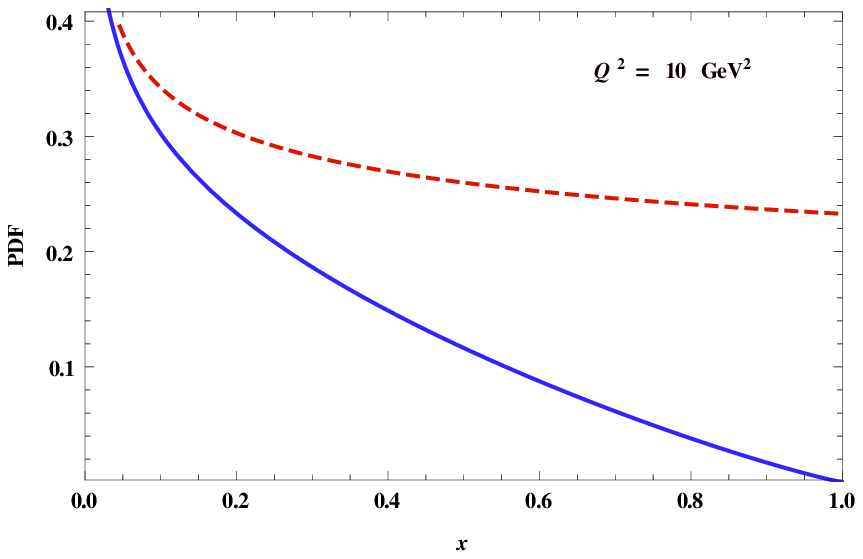}}\quad
\subfigure{\includegraphics[scale=.56]{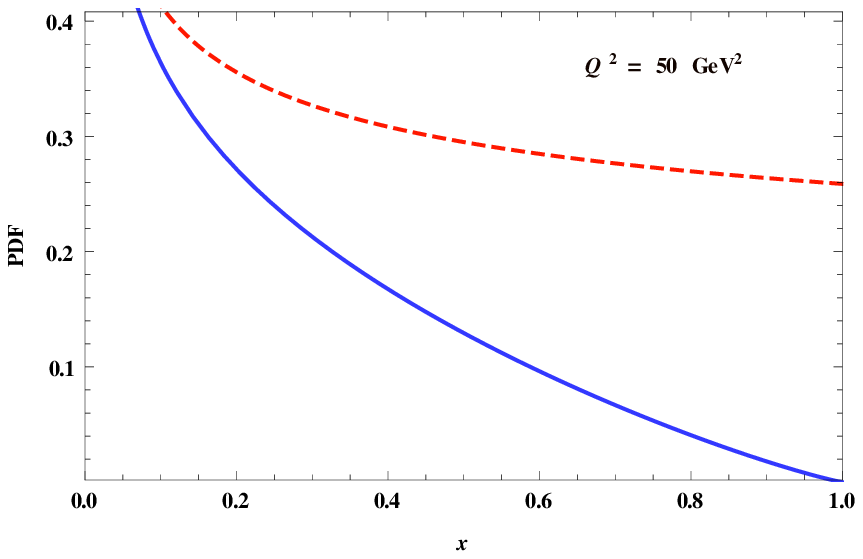}} \quad
\subfigure{\includegraphics[scale=.56]{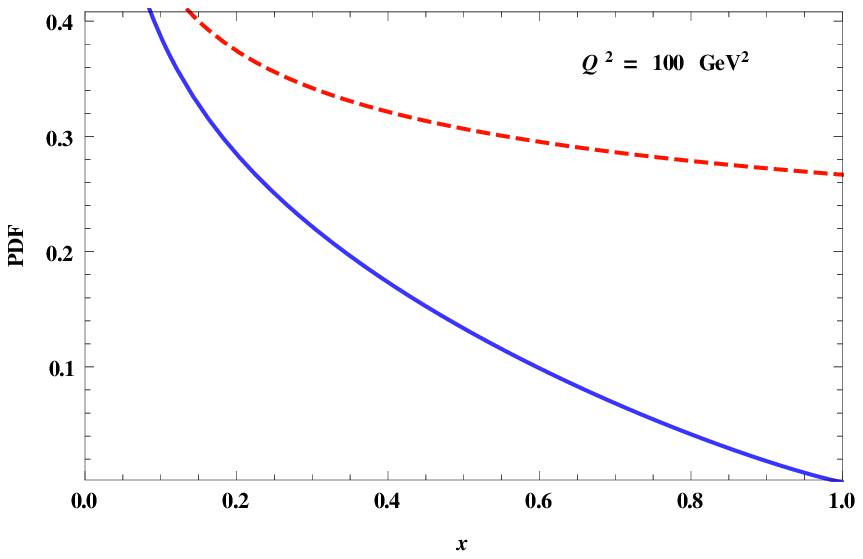}}} \quad
\caption{PDF vs $ x $}
\end{figure}

In the case of dPDF, the simplest model can be obtained if we assume $ D_{1}, D_{10} >> D_{2},...,D_{9}$. Then, the dPDF of Eq (29) is reduced to
\begin{equation}
D_{ij} \left(x_{1},x_{2},Q^{2(1)},Q^{2(2)} \right)\simeq \frac{e^{D_{0}^{ij}}}{M^{4}} \, \left( \frac{1}{x_{1}}\right)^{D_{1} \log \frac{1}{x_{2}}} \, . \, Q^{2(1)}Q^{2(2)}
\end{equation}
whereas the dPDF with the kinematic boundary (Eq (41)) is obtained as
\begin{equation}
\tilde{D}_{ij} \left(x_{1},x_{2},Q^{2(1)},Q^{2(2)} \right) \simeq \frac{e^{D_{0}^{ij}}}{M^{4}} \, \left( \frac{1}{x_{1}}\right)^{D_{1} \log \frac{1}{x_{2}}} \, .\left(\frac{1- \left(x_{1}+x_{2} \right)}{x_{1}+x_{2}} \right)^{D_{10}} . \, Q^{2(1)}Q^{2(2)}
\end{equation}
In Figure 2, we plot dPDF vs \textit{x} for a few representative values of $x_{2}=0.3, 0.1$ and $0.6$ using Eq (42) (dashed lines) and Eq (43) (solid lines).
\begin{figure}[h]
\centering
\mbox{\subfigure{\includegraphics[scale=.56]{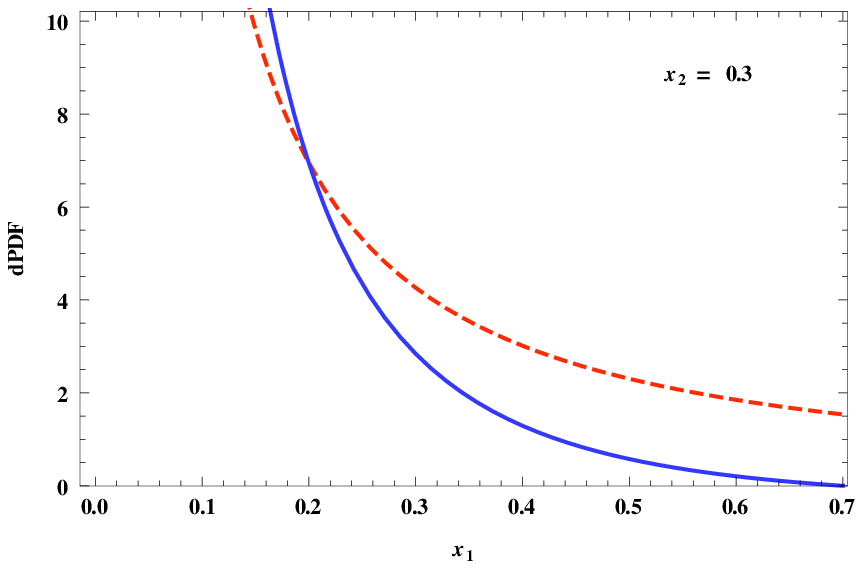}}\quad
\subfigure{\includegraphics[scale=.56]{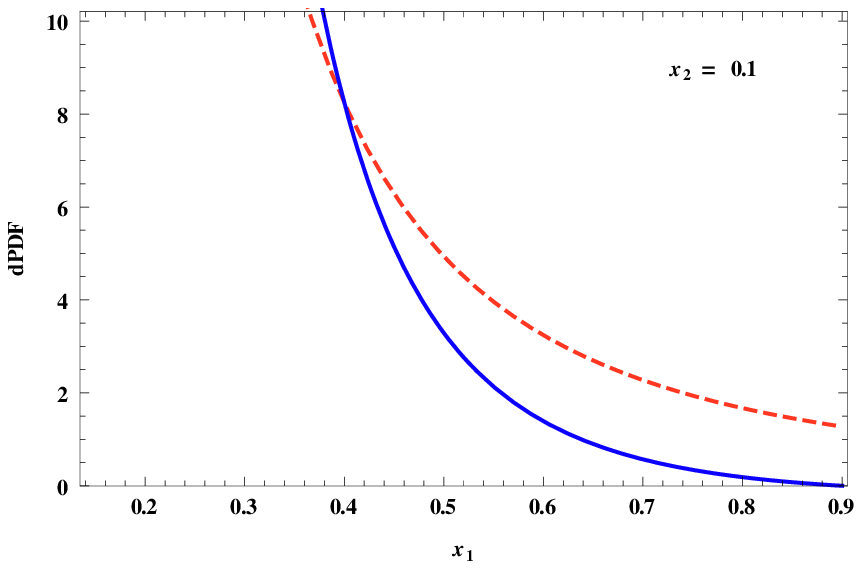}} \quad
\subfigure{\includegraphics[scale=.56]{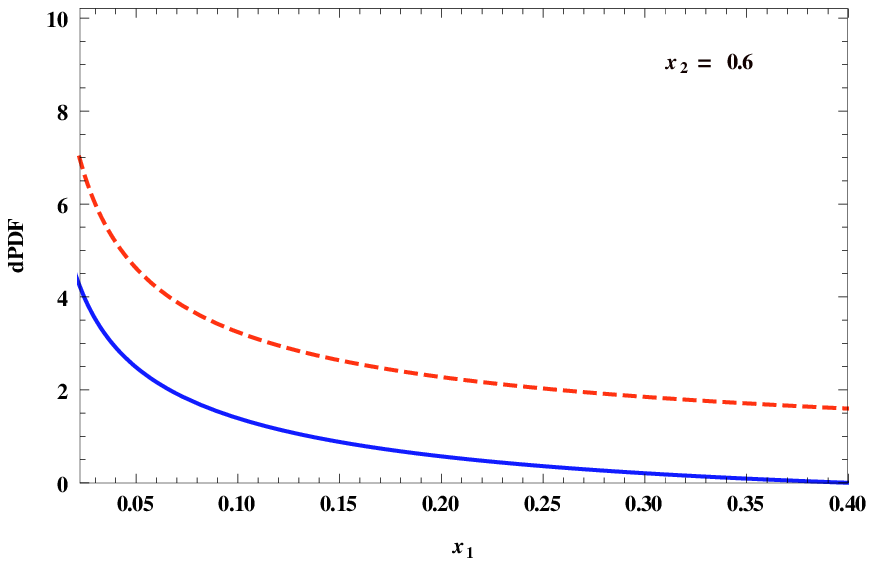}}} \quad
\caption{dPDF vs $ x $}
\end{figure}

\section{Conclusions}
In this paper, we have first extrapolated the analytical parameterization of quark densities of Ref.\cite{18} valid in the restricted kinematical region to large \textit{x} by a formal replacement of $ \frac{1}{x}$ factor to $\left(\frac{1}{x}-1 \right)$. We have then discussed the relation of the final parameterization (Eq (13)) with the standard behavior of quark densities obtained in the framework of CTEQ parameterization \cite{27}. Specific properties of self-similar parameterization are also discussed. We have then examined if the moments of the parton distribution function in QCD as obtained in Ref.\cite{28,29} are self-similar or not.\\

We have then extended the self-similar formalism of PDF to dPDF with small $x_{1},x_{2}$. We find that the constructed dPDF does not factorise into two single PDFs in conformity with QCD expectation. The dPDF has to vanish at the kinematic boundary $x_{1}+x_{2}=1$. We investigate if a smooth continuous form of dPDF can be suggested which has both the expected self-similar behaviour at small $x_{1} \, \mathrm{and} \, x_{2}$ while vanishes at the kinematic boundary. We achieve it only by introducing an additional factor $D_{10} \log \left( \frac{1}{x_{1}+x_{2}}-1 \right)$ in the defining TMD dPDF.\\

Further work is needed to investigate the feasibility of the model even though in near future, it may be difficult to expect a possibility to validate the parameterization (Eq (41)) at LHC.

\end{document}